\documentclass[conference]{IEEEtran}
\ifCLASSINFOpdf
\else
\fi
\usepackage{tikz}
\usepackage{amsmath}
\usepackage{listings}
\usepackage{xcolor}
\usepackage{enumitem}

\usepackage{filecontents}
\newcommand{\RNum}[1]{\uppercase\expandafter{\romannumeral #1\relax}}

\begin{document}
%
\title{Using SBPF to Accelerate Kernel Memory Access From Userspace}



%

\author{\IEEEauthorblockN{Boming Kong}
\IEEEauthorblockA{UC Santa Barbara\\ Email: boming\_kong@ucsb.edu} 
\and
\IEEEauthorblockN{Zhizhou Zhang}
\IEEEauthorblockA{Uber Technologies Inc.\\ Email: zzzhang@uber.com} 
\and
\IEEEauthorblockN{Jonathan Balkind}
\IEEEauthorblockA{UC Santa Barbara\\ Email: jbalkind@ucsb.edu}}



\maketitle 

\begin{abstract}
The cost of communication between the operating system kernel and user applications has long blocked improvements in software performance. Traditionally, operating systems encourage software developers to use the system call interface to transfer (or initiate transfer of) data between user applications and the kernel. This approach not only hurts performance at the software level due to memory copies between user space address spaces and kernel space address spaces, it also hurts system performance at the microarchitectural level by flushing processor pipelines and other microarchitectural state.

In this paper, we propose a new communication interface between user applications and the kernel by setting up a shared memory region between user space applications and the kernel's address space. We acknowledge the danger in breaking the golden law of user-kernel address space isolation, so we coupled a uBPF VM (user-space BPF Virtual Machine) with shared memory to control access to the kernel's memory from the user's application. In this case, user-space programs can access the shared memory under the supervision of the uBPF VM (and the kernel's blessing of its shared library) to gain non-blocking data transfer to and from the kernel's memory space. We test our implementation in several use cases and find this mechanism can bring speedups over traditional user-kernel information passing mechanisms.
\end{abstract}


%
\IEEEpeerreviewmaketitle

\section{Introduction}
Much of modern day performance engineering focuses on removing the kernel from performance critical paths.
High-performance networking and storage, for example, work to bypass the kernel and directly access hardware from user-mode~\cite{zhang2019m,dragojevic2014farm,honda2018paste,jeong2014mtcp} .
In the case where the kernel must be used, asynchronous system calls aim to allow kernel work to be done in the background while the application continues with its own critical path~\cite{Fibrilsa84:online, brown2007asynchronous}.
Where possible, optimisations may include batching system calls to reduce the number of kernel boundary crossings (for example, with PSS~\cite{PSS}).
The performance cost of system calls comes largely from kernel ABI costs and pollution of microarchitectural state, with TLB, branch predictor, and pipeline contents being modified and potentially flushed to avoid side-channel attacks.

In this paper, we propose to attack this problem by adopting one of the Linux kernel community's current favourite hammers: eBPF.
eBPF has seen enormous success for a variety of tasks inside the kernel~\cite{gregg2019bpf, eBPFSumm77:online, sharaf2022extended, zhong2022xrp, zhou2023electrode}, particularly for users who need to ship a single kernel image across a large number of systems while enabling differentiated capabilities for subsystems like network filtering.
Today, kernel users appreciate the eBPF model's ability to verify code to be run, in spite of the restrictions it places on the programming model, though the expressiveness of verifiable programs continues to grow thanks to features like BPF loops and dynamic memory allocation.

\begin{figure}[t]
    \centering
    \includegraphics[width = \columnwidth]{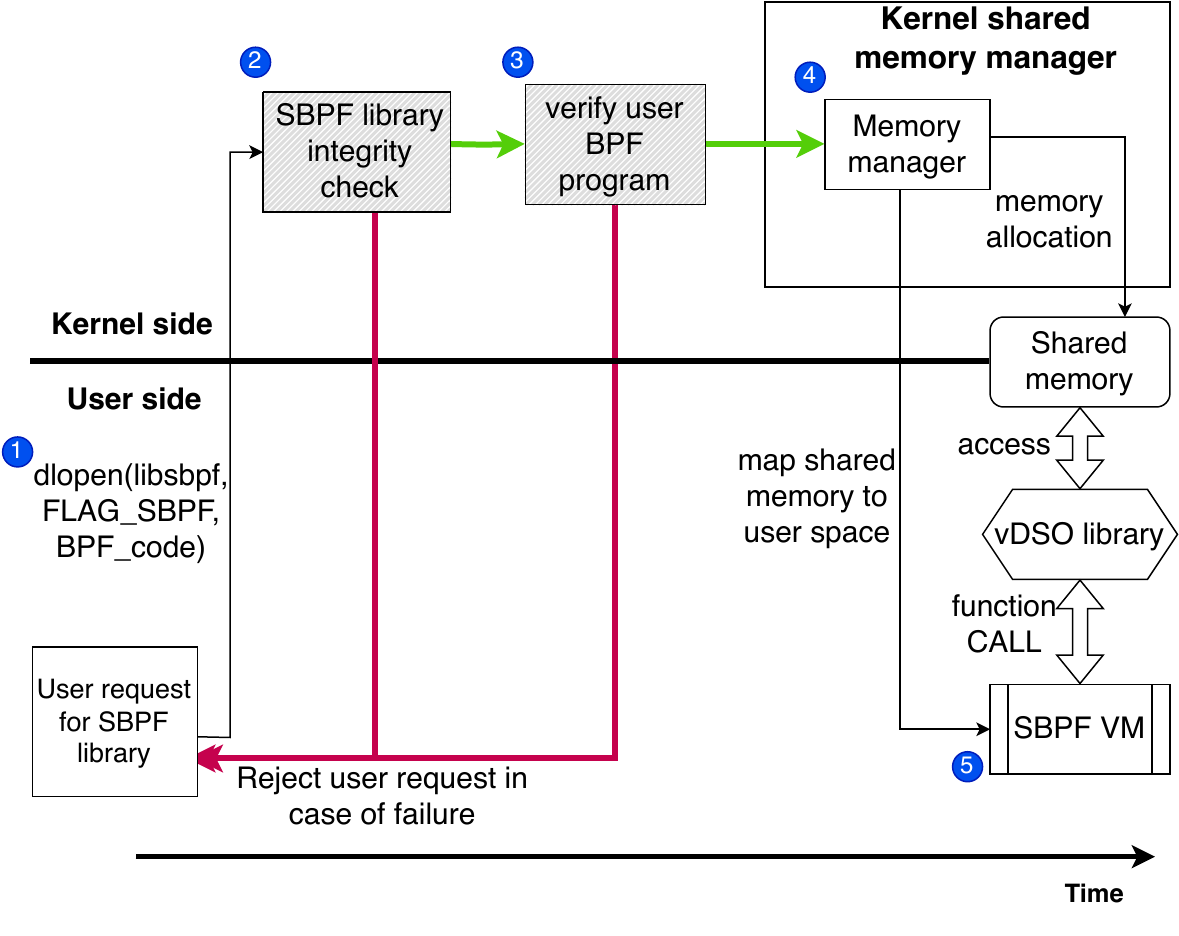}
    \caption{SBPF workflow. Red arrow means verification failed and kernel reject user request, green arrow means verification success. Shaded boxes are work in progress}
    \label{fig:sbpf-workflow}
\end{figure}

Our specific goal is to exploit eBPF in userspace (known as uBPF) to provide system features that user programs would normally require kernel assistance to provide.
Specifically, we introduce SBPF (shown in Figure~\ref{fig:sbpf-workflow}) which has direct, writable access to a small amount of kernel memory to enable fast interaction between userspace and the Linux kernel.
To load an SBPF library from a user-mode program, it first undergoes cryptographic blessing and then verification by the kernel, before being mapped as execute-only to avoid leaking kernel secrets.
While SBPF allows execution of user code in the same process as the SBPF runtime, we observe that most in-kernel uses of eBPF rely on root-equivalent privileges to load programs from user mode.
We expect that SBPF will provide extra performance for cryptographically-verified users without requiring root-level privileges and provide the user access to only the kernel memory which is needed for the user's specific task.
SBPF can also improve the performance of system calls where user-kernel data movement introduces performance overheads.

Our evaluation focuses on three workloads where writable access to kernel memory can provide meaningful performance improvements.
The first is a Prediction System Service which can update its machine learning model from user mode without relying on system calls (as required in the state of the art~\cite{PSS}), where SBPF achieves average performance improvements for macrobenchmarks of 2.92\% on aiohttp, 3.75\% on flask-blogging, 7.09\% on gunicorn, and 11.94\% on djangocms over the baseline PSS.
The second is an implementation of \texttt{copy\_to\_user} and \texttt{copy\_from\_user} which can accelerate execution of a \texttt{statfs} microbenchmark by 8\% or more for input strings shorter than 75 characters.
The last is an improvement to the BPF user ringbuffer, which is used for BPF programs that need to regularly move data between the kernel BPF program and the corresponding user process, where SBPF can accelerate over the baseline by $1.26\times$ to $5.23\times$.

Our paper makes the following contributions:
\begin{enumerate}
    \item Implementation of SBPF (based on uBPF) and its kernel-space shared memory manager 
    \item Implementation and evaluation of an SBPF-accelerated Prediction System Service
    \item Implementation and evaluation of SBPF-accelerated \texttt{copy\_to\_user} and \texttt{copy\_from\_user}
    \item Implementation and evaluation of SBPF-accelerated ringbuffer for eBPF
\end{enumerate}

\section{Background}
\subsection{System Call Overhead and Virtual Dynamic Shared Object (vDSO)}
One of the most important properties provided by modern Operating Systems is the isolation between user-space tasks and the kernel.
User-space tasks are precluded from directly accessing memory within the kernel-space.
This isolation has several benefits, including protecting kernel memory and user data, maintaining system stability against faults, and supporting better resource management.
However, an inherent challenge arises when tasks running in the user space need to access kernel-space memory.
To do that, user-space tasks must go through a system call provided by the OS.
System calls can cause a significant amount of overhead due to the need to switch between user and kernel space, save and restore registers, copy memory, handle interrupts, flush the TLB, and more.
Multiple previous studies~\cite{bagherzadeh2018analyzing, ren2019analysis} have shown that slow system calls can be the bottleneck of performant programs.

Fortunately, the Linux kernel has implemented the "virtual dynamic shared object" (vDSO) -- a small shared library provided by the kernel to accelerate the performance of certain system calls. 
The concept underlying vDSO involves the kernel's management of a compact memory block, readable and shared among all user-space programs. 
This shared memory is dynamically maintained and continuously updated by the kernel.
Several limited calls can be executed in user space without the need to be executed in kernel space.
When the call needs to read kernel memory, it can access the shared memory block instead of invoking system calls.
User-space programs are granted the privilege of reading this shared memory without worrying about the expensive overhead brought about by the context switch. 

Traditional use cases of vDSO include querying timing-related information such as \texttt{gettimeofday}, \texttt{clock\_gettime}, and \texttt{clock\_getres} or CPU-related status such as \texttt{getcpu}, \texttt{getpid}, and \texttt{datapage\_offset}.
A previous study~\cite{vdsotest} shows that calling \texttt{gettimeofday} with vDSO provides almost $4\times$ speedup compared to implementation using system calls.
Developers can also create new vDSO functions as needed, for example those used by PSS~\cite{PSS}, which we return to later.

However, an inherent constraint associated with vDSO is its restriction to read-only operations on the shared memory block from user-space programs.
This constraint ensures that the usage of vDSO still follows the rule that user space is less privileged than kernel space.
If the user-space program needs to update or write the shared data, it still needs to go through the traditional system call procedure, which impacts performance.


To address this limitation of lacking write access, we present a solution by introducing a dedicated kernel-user shared memory space per task and granting read/write privileges to that space to user-space programs.
To prevent potential memory corruption in kernel space, read/write operations within the shared memory should be executed under a user-space BPF virtual machine, the SBPF.
This virtual machine, in turn, will utilize specific vDSO functions to carry out these operations. 
Additional details of this approach are provided in Section~\ref{sec:design}.

\subsection{BPF and eBPF}
Berkeley Packet Filters (BPF)~\cite{mccanne1993bsd} were first introduced in 1993 for networking purposes.
It is a technology that allows users to specify rules to filter network packets in kernel space efficiently and effectively.
It is portable and extensible, making it suitable for different systems and architectures.
Since its creation, BPF has become an ideal tool for multiple high-performance networking applications.

Under the hood, BPF operates as a register-based virtual machine (BPF VM) embedded within the Linux kernel. 
Users create BPF programs using a restricted instruction set architecture (ISA) designed specifically for packet filtering tasks. 
After verifying the restricted programs, the kernel loads them and attaches them to particular hook points, usually associated with networking activities. 
As the network packets pass through these hook points, the BPF VM runs the relevant BPF program.


Subsequently, eBPF~\cite{eBPFIntr65:online} was introduced to extend BPF to a wider range of applications.
It offers a modular and expandable architecture that surpasses the networking-focused approach of BPF.
Compared to BPF, eBPF introduces several new program types, increases the number of registers, and supports various map types for efficient data sharing between the user space and the kernel.
It also features an ever-more expressive set of programming language features, recently including BPF loops.
eBPF also redesigns the BPF ISA to harness the inherent capabilities of contemporary processor architectures, facilitating seamless integration with the modern ISA to achieve enhanced performance.

\subsubsection{How eBPF Program Works}
The typical procedure for developing an eBPF program 
starts with writing programs in a high-level language like C.
This program is then compiled into eBPF bytecode using toolchains like LLVM or Clang.
A user-space utility tool will further load the bytecode from the user space to the kernel space.

In the kernel, the eBPF program will be verified for safety and stability, and then Just-in-Time (JIT) compiled into native machine code for efficient execution.
Once the verification and compilation are complete, the program can be attached to a specific kernel hook or event, like a network event or a system call.
When specified events occur, the eBPF program runs within the kernel context but with limited privileges for security. 
eBPF maps often come along with the eBPF program, enabling the storing and transferring of information between the kernel and user space or between different eBPF programs. 
For example, execution results can be transferred back to user space for further processing.
This workflow provides a versatile and secure way for eBPF programs to extend the capabilities of the kernel, such as network security and performance monitoring.

\subsubsection{uBPF}
User-space BPF (uBPF)~\cite{ubpf} is another key component of SBPF.
It operates entirely in user-space rather than kernel-space where traditional eBPF programs run. 
This design enables uBPF to execute eBPF programs securely without a high level of privilege.
uBPF translates BPF bytecode into native machine instructions that can be executed on the host CPU, allowing for interpretation or compilation.
It offers a simpler and more flexible way to develop and test BPF applications, as it avoids the strict security and operational restrictions of the kernel space.
It is also designed to be compatible with the eBPF ISA (used for example in the Linux kernel), so programs created with uBPF can be moved to a kernel BPF setting seamlessly.

 

\subsubsection{eBPF verification}
Verifying eBPF programs is critical to keeping the OS kernel secure and stable. 
The verification process acts as a gatekeeper, allowing only programs that will not misbehave in kernel space to execute.
It helps maintain system integrity and protect against potential security issues. 
Such a process is required to take advantage of eBPF's capabilities without compromising system security or resulting in unexpected kernel crashes.

The verifier performs several key functions, as listed below:
\begin{itemize}
    \item \textit{Safety Checks} to ensure program safety in kernel space by verifying no illegal memory accesses, potential system freezes due to loops, or other behaviors that have the potential to damage the kernel.
    \item \textit{Resource Usage Analysis} to ensure that eBPF programs do not require excessive use of CPU or memory, which can be detrimental to system performance and stability.
    \item \textit{Control Flow Analysis} to verify the validity of the program's control-flow structure and guarantee that the program ends without an infinite loop.
    \item \textit{Instruction Validation} to check the eBPF bytecode instructions for compliance with the eBPF specification, ensuring the proper instruction formats, register usage, and kernel API.
\end{itemize}

We will use the eBPF verification process in SBPF to monitor the activities of user-space programs that access shared memory, ensuring both security and optimal performance.
During the SBPF library loading process (using \texttt{dlopen()} with a new SBPF-specific flag), the kernel operates the verifier and can reject the loading of a new instance of SBPF if the verification is not met.

\section{Design of SBPF}\label{sec:design}
The SBPF VM is a specialized BPF virtual machine operating within the user-space environment to regulate read/write access to a shared memory region spanning the user-kernel boundary. 
The design of the SBPF VM focuses on facilitating a safe and efficient way to access the shared kernel-user buffer to accelerate data transfer between the kernel and the user-space program.
This virtual machine comprises three primary components: kernel-space shared memory manager, user-space SBPF VM, and SBPF library integrity verification mechanism.

The workflow for SBPF can be found in Figure~\ref{fig:sbpf-workflow}.
A user program can start using SBPF by:

(1) Loading the SBPF shared library into its own address space with \texttt{dlopen()} by using a new flag to indicate that the user program is loading an SBPF shared library, and the binary that it is attempting to load into the SBPF VM.

(2) \texttt{dlopen()} will then perform a library integrity check on the shared library sent by the user in order to make sure the library is clean and unmodified.
In the case where the integrity check fails, the kernel will reject the user's request to load the SBPF library into the user program's address space.

(3) Once the library passes the integrity check, the kernel will take the BPF binary from the integrity check module and perform a verification check on the user-space program with the kernel side BPF verifier to make sure the BPF program is safe to be executed on the shared memory.
The attempt to load the SBPF library will also be rejected if the verification check fails.

(4) Once the verification check passes, the kernel shared memory manager is responsible for allocating a per-task user-kernel shared memory space for the current user program, maintain a reference to the shared memory, and reclaiming the shared memory after the user program exits.

(5) After creating the shared memory and mapping it into the user task, the system call will return to the user program, which can then start to initialize the SBPF VM and load its BPF program into it.

We note two limitations of the present implementation.
Firstly, our implementation of the integrity check in the kernel is still in progress.
Secondly, at present the verification check for the SBPF program is performed in a limited form by uBPF in user space.


\subsection{Kernel-space Shared Memory Manager}
\begin{figure}
    \centering
    \includegraphics[width = \columnwidth]{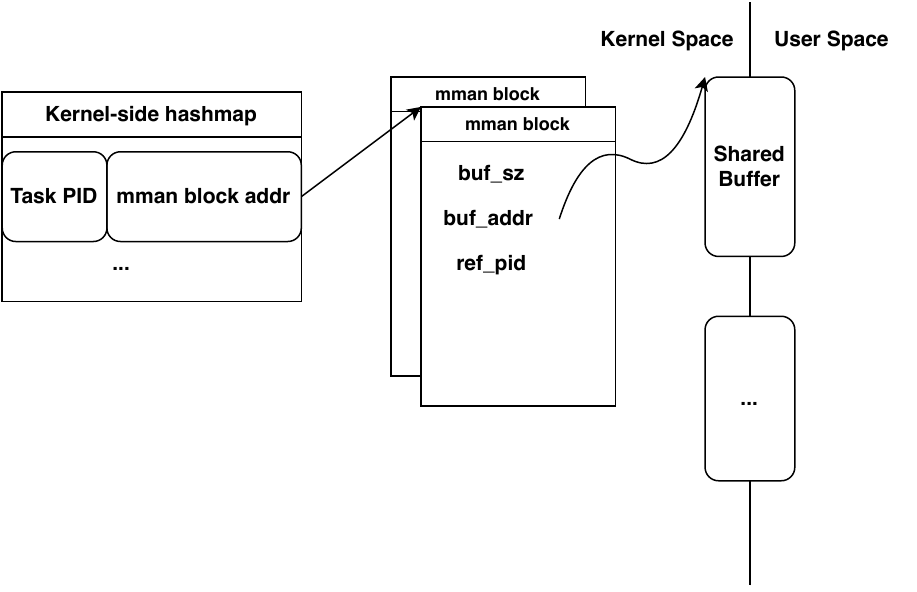}
    \caption{SBPF kernel side memory manager, ``mman block addr'' refers to the address of the memory management block}
    \label{fig:Kernel memory manager}
\end{figure}

The kernel-space memory manager, as shown in Figure~\ref{fig:Kernel memory manager}, is a hashmap-based memory pool that uses the user-space process PID to index the target memory block. Upon receiving a memory request from the user-space application from \texttt{mmap()}, the kernel-space memory manager allocates a fixed-length memory segment from the kernel-space memory heap to the user program. It also further maps this memory segment to the memory space of the user-space application.
A particular user-space application can only own the reference of one segment of the shared memory at any given time. 
Finally, when the lifecycle of the user-space program ends, the SBPF VM destructor sends an \texttt{munmap()} request to the kernel to unmap the kernel-user shared buffer. The kernel then unmaps the shared buffer from the user-space program, removes the corresponding memory registration from the kernel side hash table, and frees the shared memory.

The memory allocation process stands as a pivotal and indispensable operation within the landscape of contemporary operating systems. 
However, the efficacy of individual programs varies significantly depending on the allocation algorithm employed.
SBPF's kernel-space memory manager allows user-space programs to attach their own memory allocation algorithms and shared memory utilization techniques, effectively avoiding overheads associated with an inappropriate choice of memory management policy. 


\subsection{User-space SBPF Virtual Machine}
The SBPF virtual machine is a user-space virtual machine based on uBPF~\cite{ubpf}, which controls accesses to the kernel-user shared memory from user-space. 
The SBPF VM builds a chain of verifiable accesses to shared memory, which helps the user program to safely access shared memory without incurring the latency of the kernel context switch mechanism. The SBPF VM contains two parts -- the SBPF Just-In-Time~(JIT) compiler and the SBPF calling library.

\subsubsection{SBPF JIT Compiler}
The user-space eBPF VM~(uBPF) enables the execution of user-defined BPF programs using a Just-In-Time (JIT) compiler in user-space. To achieve this, we first compile and assemble the user program into a BPF bytecode binary file and perform a verification check using the kernel-space BPF verifier. If the check passes, the user-space BPF bytecode binary is processed by the uBPF JIT. The uBPF JIT compiler performs a seamless translation of the BPF binary file into host-ISA compatible functions, enabling efficient execution of the user-defined BPF program. This is owing to the similarity of the BPF ISA to modern ISAs such as ARM and x86-64. 


The binary translation of BPF bytecode to host-ISA compatible instructions can be affected by the incoherence of memory models between the two. Since there is no specification for a memory model for a BPF program, we chose the strict memory model provided by the x86 ISA. 


\subsubsection{SBPF calling library in vDSO}
Within the context of the BPF ISA,
the operand of the \texttt{CALL} instruction is the target sub-procedure ID number. This is unlike the \texttt{CALL} instruction in many conventional ISAs, 
which actually take target memory addresses as the operand. Consequently, performing the \texttt{CALL} instructions during runtime requires the BPF virtual machine to use a calling library. 

In our SBPF implementation, we integrated a calling library inside the vDSO, thus facilitating access to essential kernel-user shared memory functions, such as read/write operations and vector read/write capabilities.
As user-space programs lack write privileges on the vDSO memory segment, this confinement ensures the integrity of the vDSO library and prevents any injection of malicious code into the vDSO calling library.
Using some specially designed vDSO functions as the calling library for the SBPF VM, we establish a robust, verifiable, and error-free access chain to the shared memory region between the kernel and the user space.

\subsection{SBPF Library Integrity Verification Mechanism}
The user-space SBPF VM and kernel-space memory allocator have the potential to lead to security concerns.
For instance, we may worry that
attackers could load a custom SBPF library with a malicious design on the VM environment to trigger an unintended action for a custom BPF program on shared memory later. 
To mitigate this risk, we propose a mechanism for verifying the integrity of the SBPF library, as shown in Figure~\ref{fig:sbpf-workflow}. 

The SBPF library integrity verification mechanism relies on the compilation of the SBPF library with a kernel-trusted security certificate.
During compilation, the unmodified SBPF library incorporates a signature registered within the kernel (e.g. one derived from a certificate with which the kernel itself was signed).
As a result, when the user requests to load an SBPF library into its own memory space using \texttt{dlopen()}, the signature of that library will be received by the kernel side integrity check module (shown in Figure~\ref{fig:sbpf-workflow}) and sent for signature validation.
The integrity check module will allow the registration process to continue only if the signature verification is successful.
Otherwise, it means that the custom shared library that the user is trying to load is a modified version of the SBPF library, and \texttt{dlopen()} will return an invalid value and will not succeed.

Another security concern arises if 
the user program attempts to access memory beyond the boundaries of the user-kernel shared memory region.
The memory and context isolation mechanism provided by modern operating systems prevents any unauthorized access to shared memory from triggering any serious problems.
However, we need another data protection layer to ensure that access to shared memory remains exclusive to the SBPF VM, reinforcing the credibility and verifiability of memory access transactions.
To solve this problem, we apply the address space layout randomization (ASLR) technique to the shared memory.
Since the kernel memory allocator can only return the correct offset once the signature check is successful, other code will not know where the memory region starts.
This is not an absolute protection, but we note that eBPF is being regularly used to inject code into the kernel from privileged users, turning part of its security expectation into one of administrative access control.
By limiting the use of our customized \texttt{dlopen()} to applications with an SBPF-specific Linux capability, we can similarly rely on administrative access control mechanisms to only provide SBPF to relatively trustworthy applications, akin to eBPF injection into the kernel.



\section{Use cases}\label{sec:usecase}
In this section, we demonstrate the techniques for performance acceleration of three different use cases using SBPF: Prediction System Service~(PSS), \texttt{copy\_from\_user()} and \texttt{copy\_to\_user()}, and BPF user ring buffer. 


\subsection{Accelerating Prediction System Service}
Zhang et al. propose a novel low-latency Prediction System Service~(PSS)~\cite{PSS} to improve enable a learning-based replacement for heuristic-based performance tuning in both user applications and the kernel.
Using feedback-directed learning, PSS enables programmers to focus on identifying new optimization opportunities without applying extensive domain-specific knowledge or heuristic rules.
It provides fast yet accurate enough predictions, streamlining the optimization process across various runtime tasks.
Once the prediction results (for example, the observed performance of the chosen path based on the prediction) are available, the model can be rewarded or penalized accordingly.
PSS is built with a lightweight hash-based perceptron model and resides in kernel space for better reuse between multiple applications.
To reduce the overhead of system calls when user space programs need to invoke \texttt{predict()}, PSS shares the perceptron model with user-space programs through vDSO for faster reads. 
However, the efficacy of user-space model updates is impeded by the considerable overhead incurred during system calls, leading the authors to employ a batching technique to lower the cost.
This batching results in a less fresh model, potentially decreasing performance due to poorer prediction results.

\begin{figure}[t!]
    \centering
    \includegraphics[width = \columnwidth]{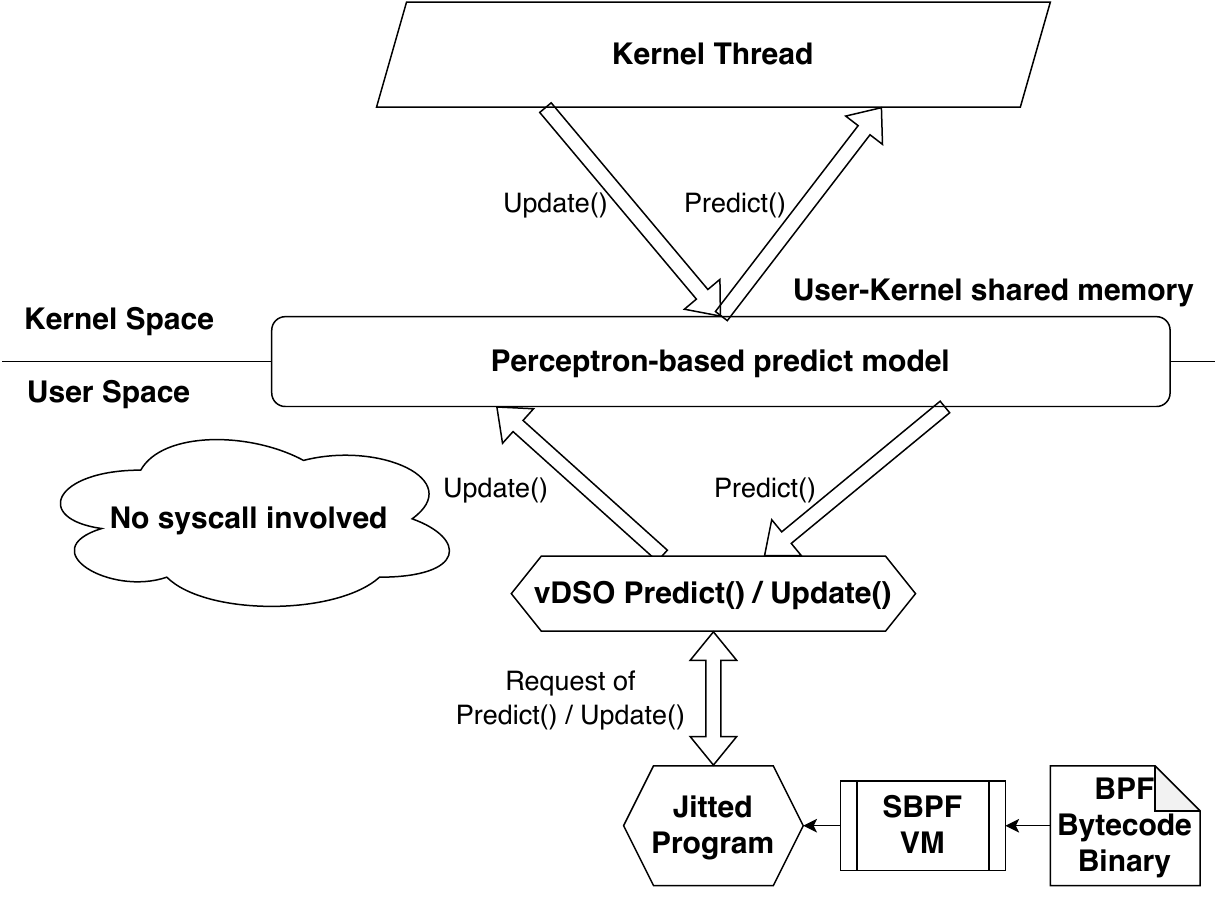}
    \caption{Accelerating PSS by deploying its prediction model directly in user-writable shared memory}
    \label{fig:PSS accel}
\end{figure}

Utilizing SBPF, we address this issue by deploying the entire perceptron-based machine learning model within the user-kernel shared memory, facilitating seamless interaction between user-space and kernel-space components. 
As shown in Figure~\ref{fig:PSS accel}, the access to this model, achieved through the \texttt{predict()} and \texttt{update()} functionalities, can be verified and managed via the uBPF JIT (Just-In-Time) compiler. 
This integration significantly improves the performance and efficiency of PSS, paving the way for improved usability and practicality in diverse real-world applications.

\subsection{Accelerating \texttt{copy\_from/to\_user()}}\label{sec:copyuser}
The Linux kernel-space APIs, \texttt{copy\_from\_user()} and \texttt{copy\_to\_user()}, hold significant prominence across diverse system calls due to their essential role in ensuring secure data transfer between user and kernel spaces within the Linux environment. When utilizing these two Linux APIs, the user provides the user-space virtual memory address, and these two functions check the permissions associated with the source and destination addresses. Further, the user-space virtual address is translated to the user-space physical address using the user-space page table, 
and the internal memory copy operation is started.
However, the inherent memory copy operations within these APIs can be both time-consuming and energy-inefficient. To mitigate this problem, we utilize user-kernel shared memory to perform the data transfer and introduce two new Linux kernel-side APIs, \texttt{sbpf\_copy\_from\_user()} and \texttt{sbpf\_copy\_to\_user()} to allow the kernel code to read user passed arguments and write system call results within the shared memory. \texttt{sbpf\_copy\_from\_user()} and \texttt{sbpf\_copy\_to\_user()} only accept memory offsets within the shared memory, so there is no need to do complex permission checks. Nor do we need expensive memory copy operations every time the user program needs to send some data to the kernel as both sides have the shared memory already mapped.

\begin{figure}[t!]
    \centering
    \includegraphics[width = \columnwidth]{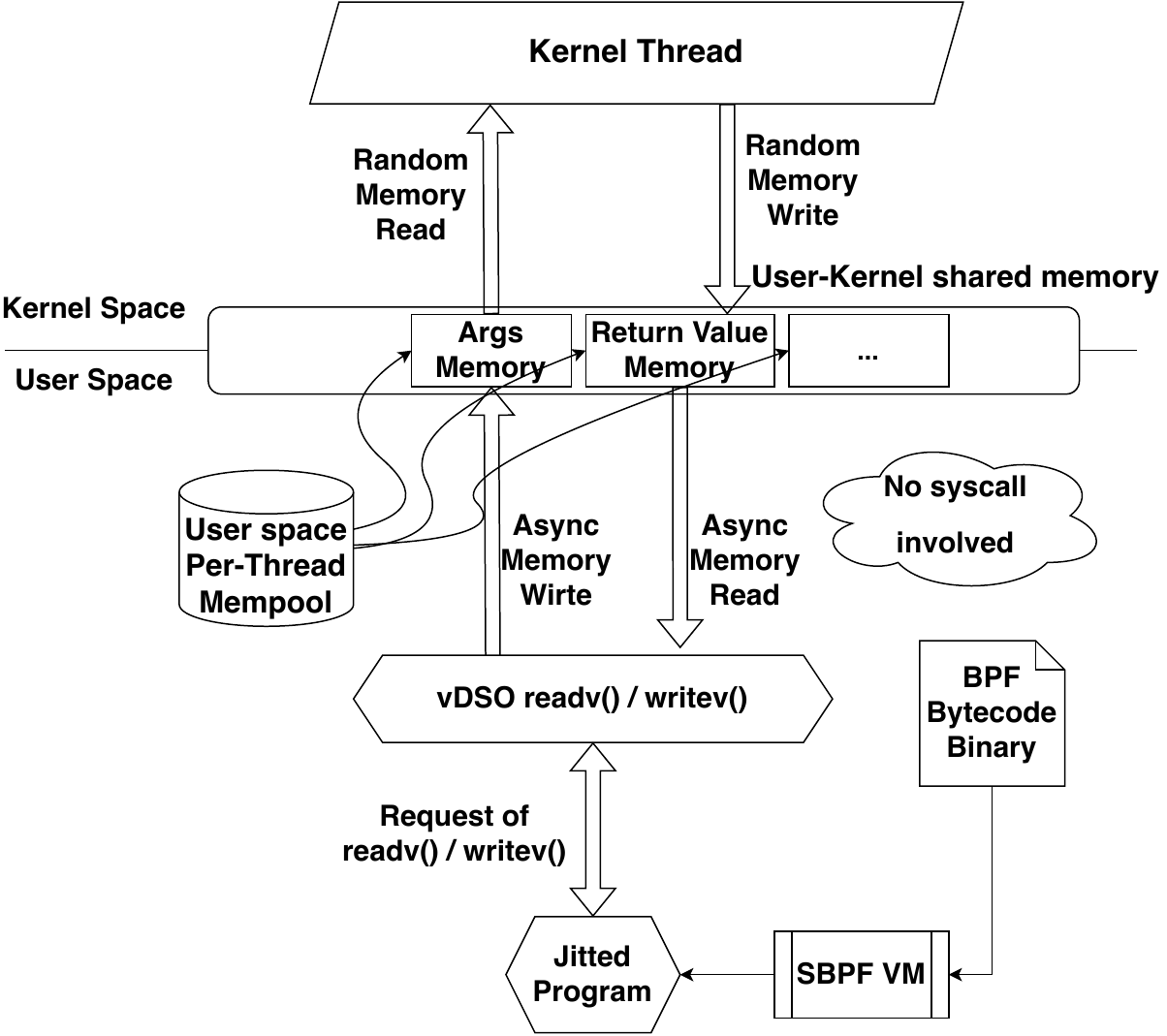}
    \caption{Accelerating \texttt{copy\_from\_user()} and \texttt{copy\_to\_user()} by deploying a user-space per-thread memory pool on shared memory to achieve zero-copy data communication between user-space program and kernel}
    \label{fig:copy_to/from_user() accel}
\end{figure} 

In order to ensure data isolation among different threads within a task, SBPF introduces a lightweight per-thread memory pool associated with each task's SBPF shared memory (shown in Figure~\ref{fig:copy_to/from_user() accel}). Each thread is granted access to two fixed-length memory blocks from the user-space per-thread memory pool during its entire life cycle. These two memory blocks are "args memory" and "return value memory". When user threads need to transfer some data to kernel space, they first need a BPF program to call the vDSO library functions to send their data to args memory. After that, the user program needs to submit the BPF program to the SBPF VM for verification. All programs that tend to access memory beyond their two allowed memory blocks will be rejected. Once the verification passes, the jitted program will perform its job of writing all data into the args memory. After all the data has been sent to args memory successfully, the user thread can invoke the desired kernel system call and provide its own thread ID as a system call argument. When the kernel has been invoked by the user space system call, there is no need to use the \texttt{copy\_from\_user()} or \texttt{copy\_to\_user()} functions to perform a memory copy between the userspace memory and kernel memory. Instead, the kernel only needs to call \texttt{sbpf\_copy\_from\_user()} and \texttt{sbpf\_copy\_to\_user()}, and these two APIs will calculate the address of args memory based on the thread ID provided by the user thread and return the result to the rest of the system call. Since the APIs we propose do not need to perform memory copies, they can free user programs from the high latency brought by frequent memory copy operations.


\begin{lstlisting}[float, caption={Traditional eBPF program to drain one data element with fixed size from the user ringbuffer. user\_ringbuf\_drain() is an API provided by eBPF, and users need to call it first to fetch data for custom callback routine (drainer() here).}, label={lst:ring-buff}]
long drainer(struct bpf_dynptr *ptr){
  int *data = bpf_dynptr_data(ptr, sizeof(int));
  if(!data){
    bpf_printk("Failed to get data");
    return 1;
  }
  ...
  return 0;
}
int bpf_main(){
  int err = user_ringbuf_drain(&URING, drainer);
  ...
}
\end{lstlisting}

\subsection{Accelerating the BPF user ring buffer}
\begin{figure}[t!]
    \centering
    \includegraphics[width = \columnwidth]{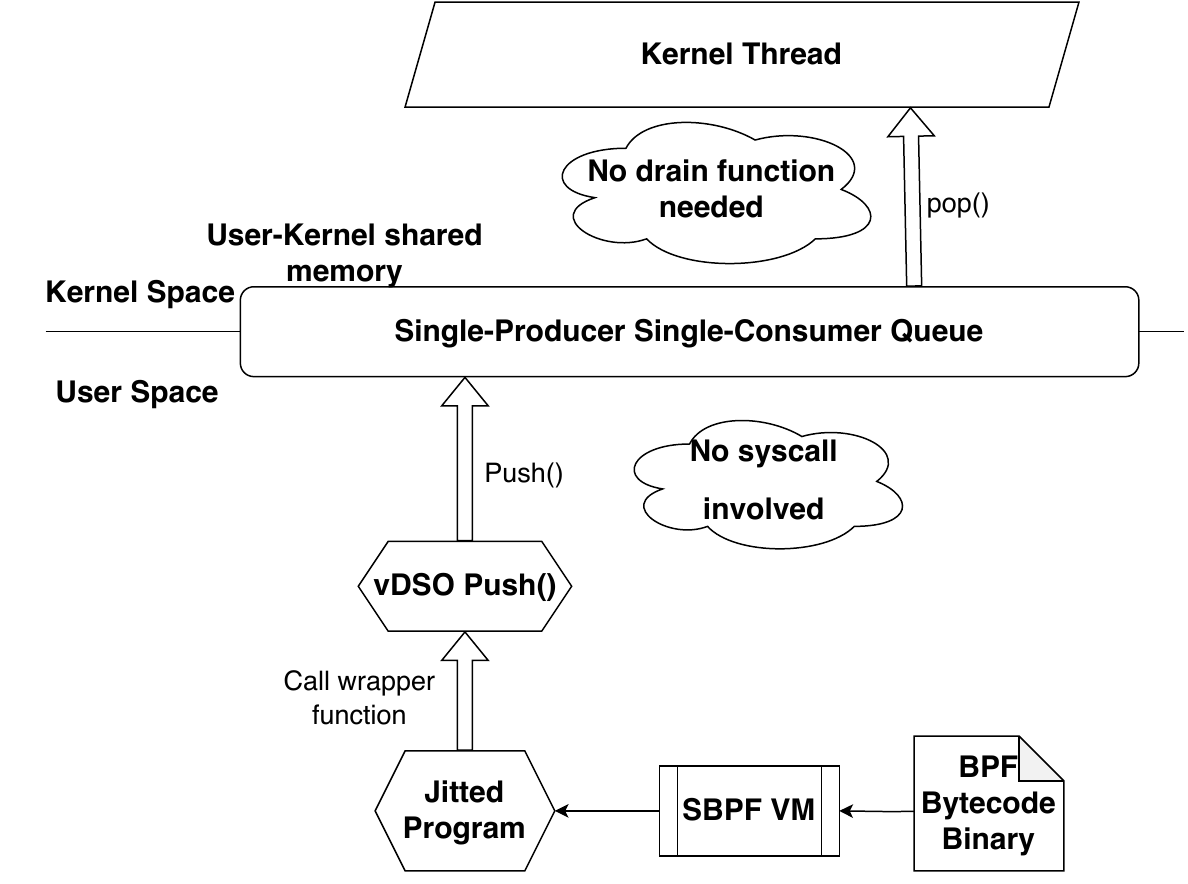}
    \caption{Accelerating the BPF user ring buffer by deploying the SPSC ring buffer inside kernel-user shared memory provided via SBPF}
    \label{fig:SPSC accel}
\end{figure}

The BPF user ring buffer introduces a novel queue type that enables eBPF programs to receive data from userspace. When a userspace program needs to push some data into the user ringbuffer, it needs to submit the data into the \texttt{bpf()} system call. Once the attached BPF program on the kernel side has been invoked, it must call a drain function to pop one piece of fixed-length data from the user ring buffer (Figure~\ref{fig:SPSC accel}). The latency brought by invoking the system call and drain function is high, and the fixed-length constraints make the current eBPF program less flexible. SBPF offers an optimized implementation of this user ring buffer by transforming the entire kernel-user shared memory into a Single-Producer Single-Consumer (SPSC) queue (Figure~\ref{fig:SPSC accel}). The user side push() function is implemented with a vDSO and the SBPF VM provides a wrapper function for user BPF programs. In this situation, the kernel can directly retrieve flexible-size data from the SPSC queue, including in batches, without suffering the latency and limitations brought by any system call and drain functions. This optimization results in a more efficient data transfer mechanism between kernel threads and user-space processes.

\section{Evaluation}
We evaluated SBPF on an 8 core Intel Core i9-9900KF processor with 3.60GHz frequency.
The processor has 256 KiB L1 instruction cache, as well as 256 KiB L1 data cache, 2MiB L2 cache for each core, and one 16 MiB L3 cache that is shared among all cores. 
The memory on our test machine is 32GB, and the operating system kernel that handles the tests is Linux 6.2.10.

We evaluated the three use cases presented in Section~\ref{sec:usecase}.
First, we evaluate the python-macrobenchmark~\cite{python-macrobenchmark} to illustrate speedup over PSS~\cite{PSS}. 
Second, we show the acceleration of \texttt{copy\_from\_user()} and \texttt{copy\_to\_user()} by examining the performance of the SBPF optimized \texttt{sbpf-statfs()} system call compared to the native \texttt{statfs()} system call on Linux~\cite{statfs2L80:online}. 
Finally, we compare the performance difference between the SBPF optimized BPF user ringbuffer and the native Linux BPF user ringbuffer.

\begin{figure*}[t!]
    \centering
    \includegraphics[width = 0.5\textwidth]{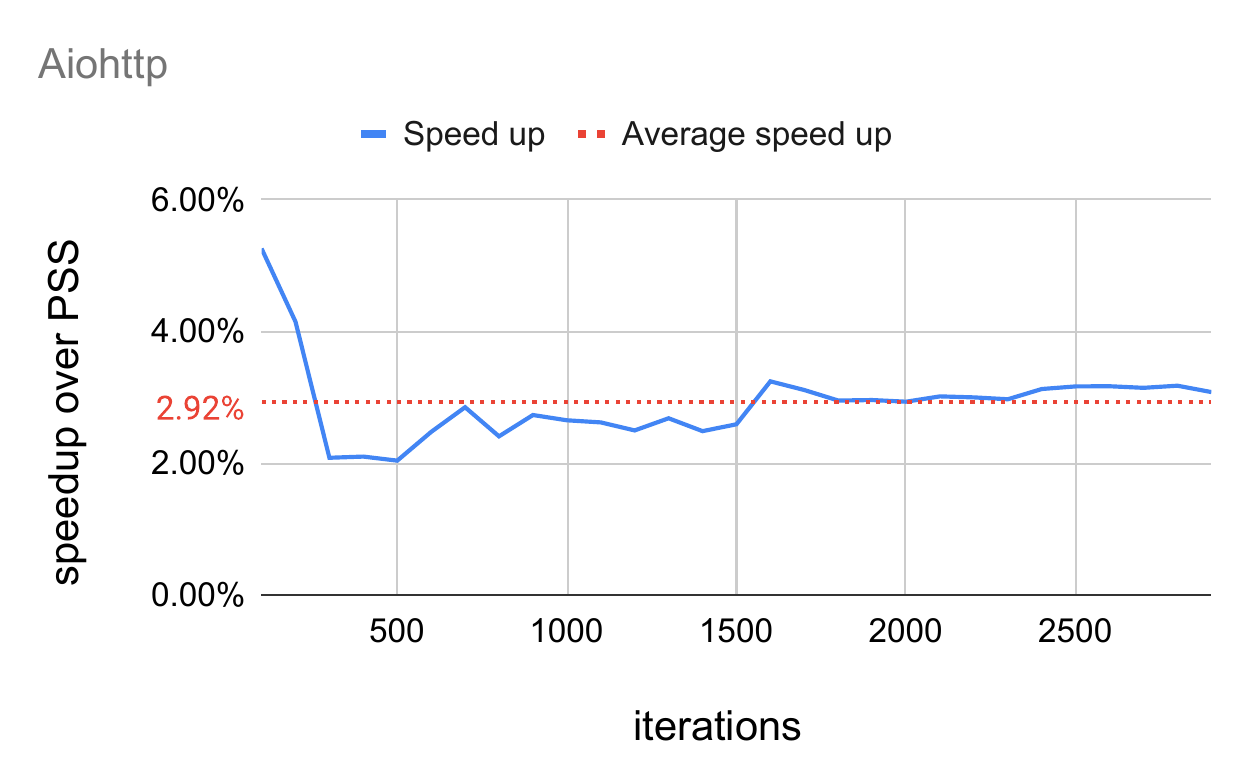}\hfill
    \includegraphics[width = 0.5\textwidth]{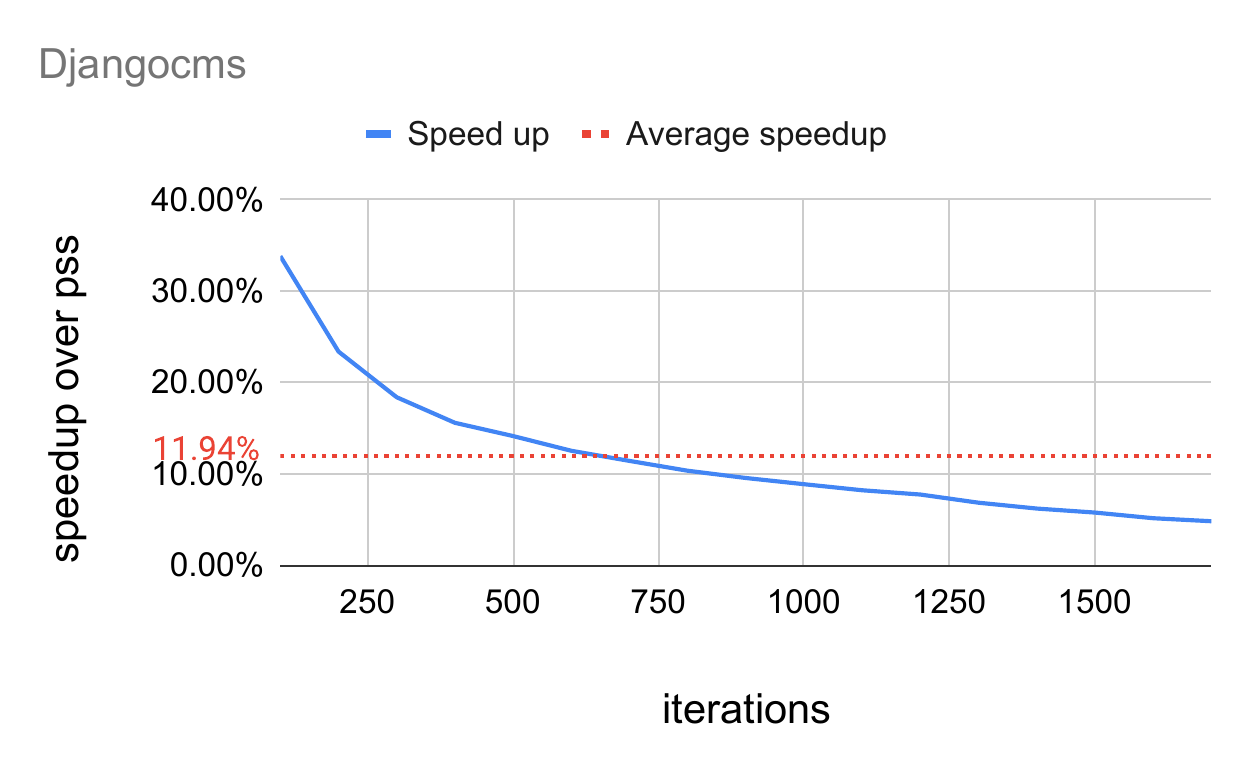}\par
    \includegraphics[width = 0.5\textwidth]{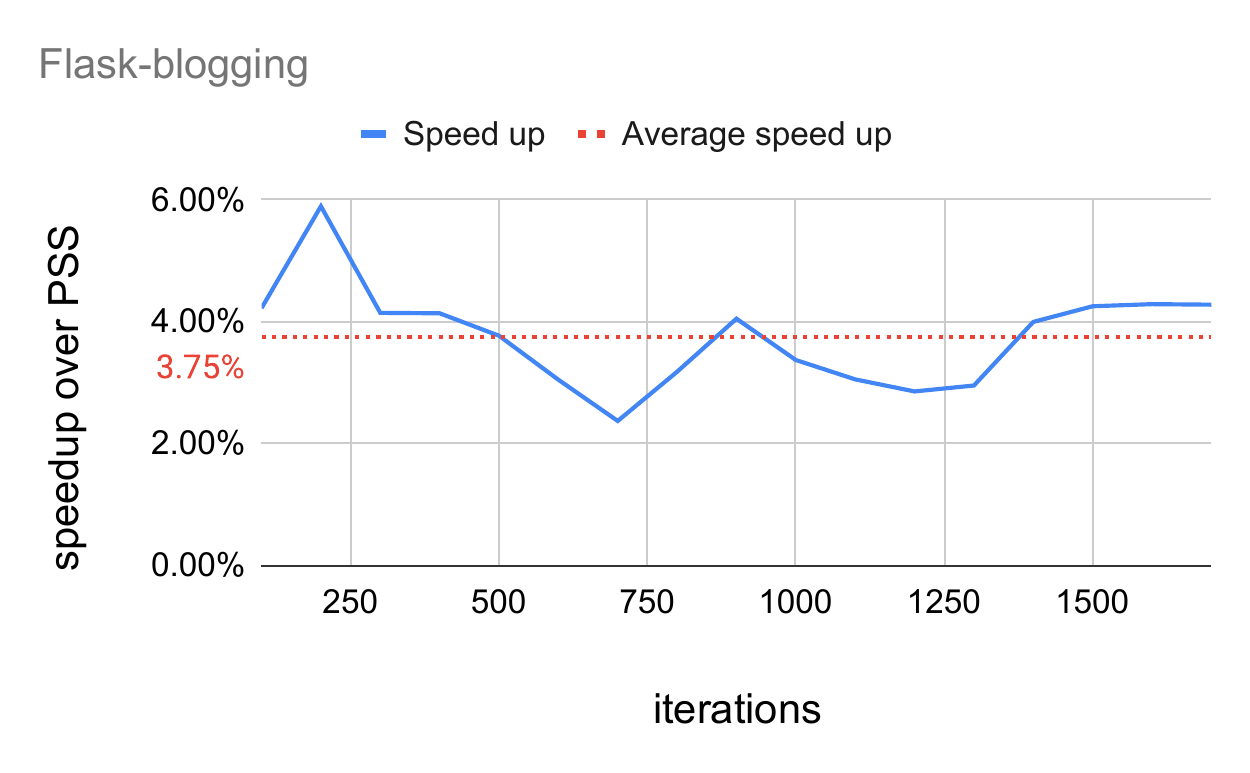}\hfill
    \includegraphics[width = 0.5\textwidth]{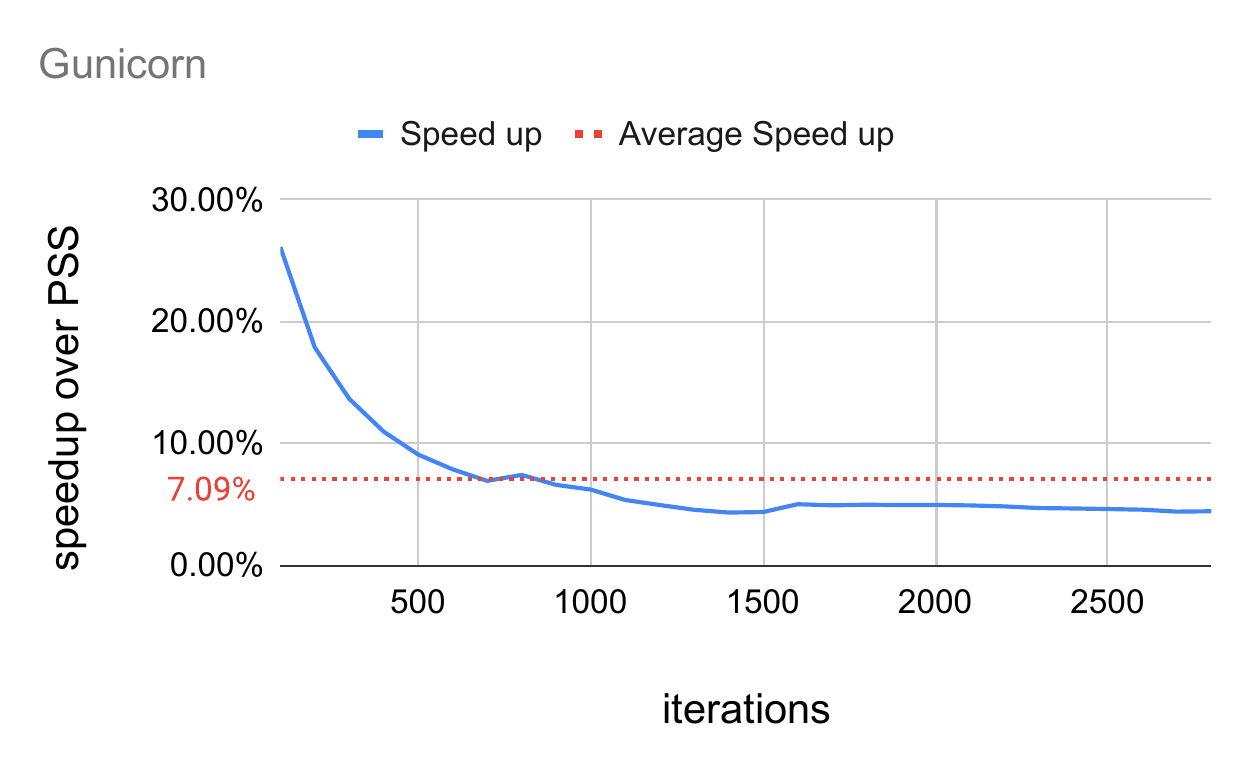}
    \caption{Result of python-macrobenchmark}
    \label{fig:macro-result}
\end{figure*}

\subsection{Python-macrobenchmark results}
There are three sets of macrobenchmarks included in the original PSS paper: hardware lock elision (HLE), memory management subsystem, and PyPy JIT parameter tuning.
We choose the PyPy parameter tuning as a use case in this paper to show how we utilize SBPF to improve the performance of PSS.
In the original work, PSS was able to improve performance of PyPy JIT parameter tuning versus the off-the-shelf PyPy's own parameter tuning.
We compare our performance to the already PSS-optimised version of PyPy.

The tests are configured with PyPy 7.3.3~\cite{PyPy81:online} as the JIT compiler. 
We choose python-macrobenchmark~\cite{python-macrobenchmark} as the workload, which is also used in the original PSS evaluation.
Python-macrobenchmark contains many popular real-world applications for user level. 
For our evaluation, we selected four applications from python-macrobenchmark: Aiohttp~\cite{aiohttp}, Flask-blogging~\cite{flask}, Gunicorn~\cite{gunicorn}, and Djangocms~\cite{djangocms}, to use as macrobenchmark test cases for SBPF-optimized PSS.
Similarly to the evaluation in PSS, we perform 1900 iterations for aiohttp, 1700 iterations for flask-blogging, 2800 iterations for gunicorn, and 1700 iterations for djangocms as the workload.
In addition, each of our tests is executed 4 times, and we report the average number as a result. 
Initially, we set up PSS in our test environment and recorded its performance metrics, labeling the data as the baseline result. 
Next, we evaluated the time consumption for the SBPF-optimized PSS as optimized results. 
We then compared these optimized results with the baseline to calculate the percentage of performance improvement, shown in Figure~\ref{fig:macro-result}.

The plot clearly shows that the SBPF-optimized PSS outperforms the original PSS in all cases on four types of macrobenchmarks. 
As depicted in the figure, the SBPF-optimized PSS achieves average performance improvements of 2.92\% on aiohttp, 3.75\% on flask-blogging, 7.09\% on gunicorn, and 11.94\% on djangocms over the baseline PSS. 
Consistent improvement in performance in the test curves indicates that combining SBPF optimization with PSS leads to a noticeable improvement in performance due to fresher prediction state for the PSS-held model weights.

\begin{figure}[t!]
    \centering
    \includegraphics[height = 6cm, width = \columnwidth]{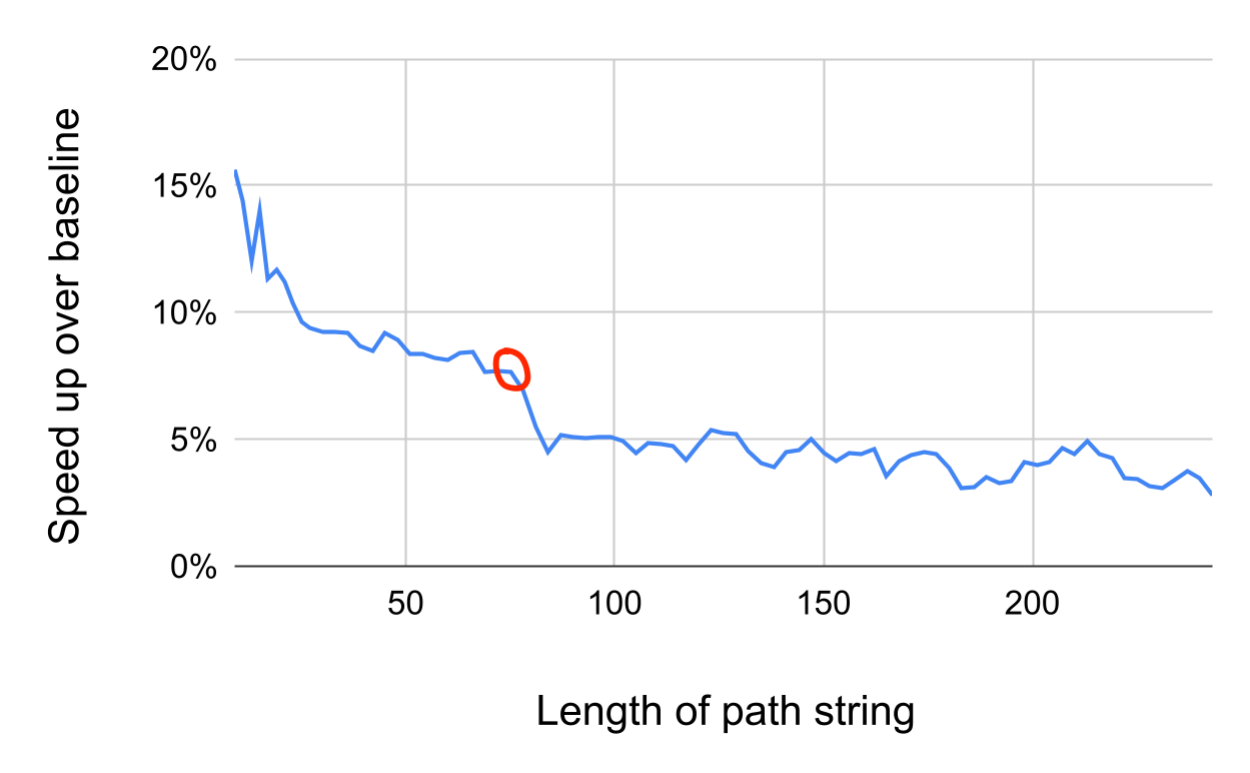}
    \caption{SBPF speed up against baseline system call for \texttt{statfs()}}
    \label{fig:sbpf-syscall}
\end{figure}

\subsection{Linux system call result}
In this test, we introduce a new system call named \texttt{sbpf-statfs()}, which replicates the functionality of the original \texttt{statfs()} system call in Linux 6.2.10. 
The only divergence in functionality is that the original \texttt{statfs()} uses \texttt{copy\_from\_user()} and \texttt{copy\_to\_user()} to communicate with the user space program, while \texttt{sbpf-statfs()} uses the new kernel APIs, \texttt{sbpf\_copy\_from\_user()} and \texttt{sbpf\_copy\_to\_user()} that we implemented in Section~\ref{sec:copyuser} to replace all the \texttt{copy\_from\_user()} and \texttt{copy\_to\_user()} calls that exist in \texttt{statfs()}. 
\texttt{sbpf\_copy\_from\_user()} calculates the beginning address of the user program's ``args memory'' and returns the result to the rest of the system call when it needs to read from user arguments, and \texttt{sbpf\_copy\_to\_user()} calculates the beginning address of the user program's ``return value memory'' and returns the result to the rest of the system call when it needs to write the result.




Our experimental setup involved the deployment of a test program on a Linux kernel version 6.2.10, patched to support the \texttt{sbpf-statfs()} function. 
This program tested the \texttt{statfs()} system call 10,000 times with different valid file paths of lengths from 9 to 243 characters to measure the amount of time it took.
These measurements served as our baseline data. 
We then executed the same workload using the \texttt{sbpf-statfs()} system call and recorded its time consumption to measure the performance difference.
For robustness, each test was run seven times, with the highest and lowest values discarded; the final result was derived by averaging the remaining values.
The final results were derived by calculating the speedup of \texttt{sbpf-statfs()} time consumption relative to the baseline of the original system call.
The results of this experiment are illustrated in Figure~\ref{fig:sbpf-syscall}.

According to the result of Figure~\ref{fig:sbpf-syscall}, the performance acceleration remains over 8\% when the length of the input string is less than 75 characters (marked in Figure~\ref{fig:sbpf-syscall} by a red circle). 
After this point, the speed up scales down to 4\% and becomes relatively stable. Through the curve on the figure, we can say that by using \texttt{sbpf\_copy\_from\_user()} and \texttt{sbpf\_copy\_to\_user()} to replace current \texttt{copy\_from\_user()} and \texttt{copy\_to\_user()} Linux kernel API, we can have a 4\% or higher speedup against the traditional \texttt{statfs()} implementation in Linux 6.2.10 for common path string lengths. 

\begin{figure}[t!]
    \centering
    \includegraphics[height = 6cm, width = \columnwidth]{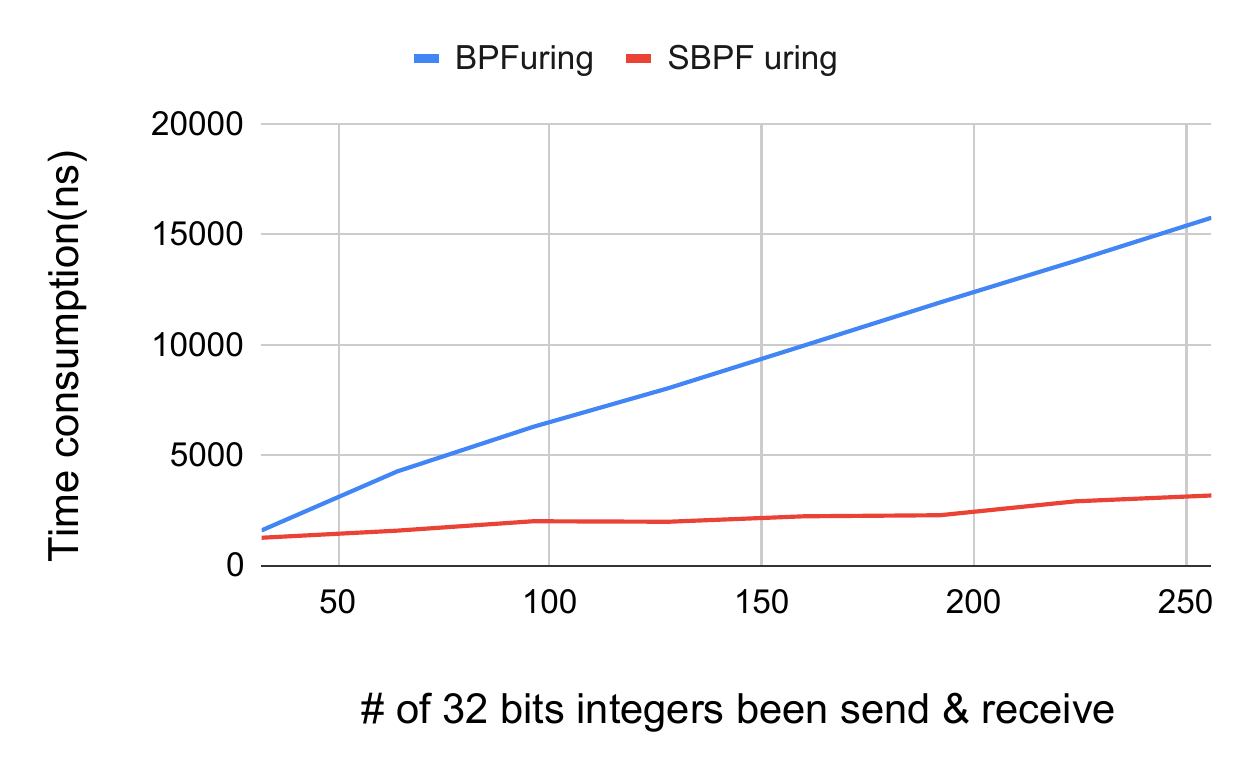}
    \caption{SBPF speedup against baseline system call}
    \label{fig:sbpf-spsc-result}\end{figure}

\subsection{SBPF User Ringbuffer result}
While eBPF is being heavily adopted in industry, particularly for cloud workloads, many of the use cases for brand new Linux kernel eBPF features are proprietary, making real-world application benchmarks difficult to find.
Unfortunately, at time of writing we were unable to find any real-world applications that made use of the BPF user ringbuffer, which is a relatively new kernel feature.
Instead, we created a synthetic microbenchmark to measure the effectiveness of the SBPF user ringbuffer.
Specifically, we developed a microbenchmark that logs the timing overhead when sending data to shared ring buffer plus the overhead of kernel space eBPF program to consume all the data from the user-kernel shared ring buffer. 
We set up this experiment on the conventional eBPF user ringbuffer and mark the result as the baseline. After that, we measure the overhead for our SBPF-optimized user ringbuffer in the same way and perform a comparison between these two sets of data to draw our conclusion on this test case.

The amount of data moved by the ringbuffer varies from 32 integers to 256 integers, the increment for each test case falls in a uniform distribution of 32 each. We measure the overhead expenditure by both the consumer and the producer for those data in nanoseconds.
Note that we conduct distinct measurements for both the baseline and SBPF ringbuffers, individually on the kernel-side consumer and user-side producer, and sum them up as the final score for each group of tests. This approach ensures that the outcomes remain unaffected by Linux scheduler and other architecture related context switch mechanisms. We carry out five executions for each of the cases, eliminate the shortest and longest time duration, and then calculate the average time consumption as the final result. Figure~\ref{fig:sbpf-spsc-result} presents our result.

In Figure~\ref{fig:sbpf-spsc-result}, we can find that the timing overhead for the conventional eBPF user ringbuffer exhibits a pronounced escalation with increasing workload. This phenomenon may be attributed to the need for utilization of the drain function along with dynamic pointers to access shared memory in the eBPF ring buffer. 
For the case of SBPF, overhead growth is comparatively gradual compared to baseline. 
In particular, SBPF yields an acceleration of $1.26\times$ to $5.23\times$ compared to the baseline approach. In this case, we can observe the substantial performance elevation brought about by the SBPF optimization over the current eBPF user ringbuffer data structure.

\section{Related work}
\subsection{Broad Use of eBPF}
eBPF has been shown to be one of the most versatile technologies in modern Linux, handling a wide range of tasks~\cite{gregg2019bpf, eBPFSumm77:online, sharaf2022extended}.
Some example use cases include load balancing~\cite{chen2020machine}, inter-VM traffic monitoring~\cite{hong2018design}, guard DNS privacy~\cite{rivera2020leveraging}, process confinement~\cite{findlay2020bpfbox} and many more.

Recently, eBPF has also been applied to reduce the context switch overhead of certain I/O operations.
Traditionally, an external I/O request and response need to go between user-space and kernel space, causing significant context-switching overhead.
With eBPF, request processing can be offloaded to networking~\cite{ghigoff2021bmc}, storage~\cite{zhong2021bpf}, and scheduling~\cite{humphries2021ghost} subsystems.

User-mode uses of BPF include bpftime~\cite{zheng2023bpftime} which can perform functions like user-mode uprobe and system call hooks.
Performing these in user mode enables significant performance improvements over the traditional kernel components which perform the same functions.


\subsection{Accelerating System Call}
A variety of efforts have been made to speed up system calls in both academia and industry.
For example, Miemietz et al.~\cite{miemietz2022fast} propose a new framework to reduce the overhead of the system call, by adding a new layer, ``\textit{Fastcall}'', into the traditional OS architecture to provide specialized and fast access to OS functions.
Privbox~\cite{kuznetsov2022privbox} introduces a semi-privileged execution mode, which is more privileged than user-space, yet less privileged than the kernel. 
Such a mode reduces the context-switch overhead and makes system calls faster, while maintaining security by being contained within a sandboxed environment. 
FlexSC~\cite{soares2010flexsc} presents a technique to manage system calls without the use of exceptions, thus reducing the overhead typically associated with system calls.
FlexSC includes a specialized thread scheduler for system calls and a user-mode thread package, FlexSC-Threads, which translates traditional synchronous system calls into exception-less ones.
Cassyopia~\cite{rajagopalan2003cassyopia} leverages compiler optimizations to accelerate system call overhead.
Cassyopia reduces the number of address space crossings, by profiling system call behavior and applying compiler transformations based on profiles.
Another approach to optimizing system calls is kernel bypassing. 
Kernel bypassing allows an I/O-bound application to access the underlying device directly, eliminating the need to switch between the kernel and the user space, thus improving performance in both networking~\cite{HomeDPDK63:online,firestone2018azure, sadok2021we, xing2022bedrock} and storage devices~\cite{kim2016nvmedirect, yang2017spdk, kwon2020fvm, StorageP60:online}.
Starting with Linux kernel 5.1, \textit{io\_uring}~\cite{iouringp16:online} was created to facilitate efficient I/O activities.
It is able to support both asynchronous and synchronous I/O operations efficiently and effectively.

AnyCall\cite{anycall} proposed a new method to use an in-kernel BPF bytecode executor as the middle layer to decouple the user-kernel transition during the system call. This approach reveals eBPF's potential as a proficient and authenticated ISA standing between user-space and kernel-space to diminish the latency during the processor privilege transition.

\section{Conclusion}
Performance improvements for applications that run on modern operating systems regularly hit the wall of expensive unexpected context switches between processors' privileged mode and unprivileged mode. SBPF solves this by proposing a new communication platform for both the user and the kernel to perform data transfer in a verifiable and more efficient way than today's Linux supports. 

Based on the results we obtained from our three use cases, we can observe significant and stable performance improvements across the board. Although some SBPF features are still in progress, we believe that the utilization of SBPF can help current operating systems, as well as many user applications, to perform better in a variety of scenarios which require bi-directional user-kernel data transfers.






%




\bibliographystyle{IEEEtran}
\bibliography{references}

\end{document}